\documentclass[preprint,showpacs,preprintnumbers,amsmath,amssymb]{revtex4}

\usepackage[dvipdfmx]{graphicx}
\usepackage{dcolumn}% Align table columns on decimal point
\usepackage{bm}% bold math
\usepackage{color}

\begin{document}

%\preprint{APS/123-QED}
\newif\ifplot
\plottrue
%\plotfalse
\newcommand{\RR}[1]{[#1]}
\newcommand{\intsum}{\sum \kern -15pt \int}
\newfont{\Yfont}{cmti10 scaled 2074}
\newcommand{\Y}{\hbox{{\Yfont y}\phantom.}}
\def\O{{\cal O}}
\newcommand{\bra}[1]{\left< #1 \right| }
\newcommand{\braa}[1]{\left. \left< #1 \right| \right| }
\def\Bra#1#2{{\mbox{\vphantom{$\left< #2 \right|$}}}_{#1}
\kern -2.5pt \left< #2 \right| }
\def\Braa#1#2{{\mbox{\vphantom{$\left< #2 \right|$}}}_{#1}
\kern -2.5pt \left. \left< #2 \right| \right| }
\newcommand{\ket}[1]{\left| #1 \right> }
\newcommand{\kett}[1]{\left| \left| #1 \right> \right.}
\newcommand{\scal}[2]{\left< #1 \left| \mbox{\vphantom{$\left< #1 #2 \right|$}}
\right. #2 \right> }
\def\Scal#1#2#3{{\mbox{\vphantom{$\left<#2#3\right|$}}}_{#1}
%\kern -2pt
{\left< #2 \left| \mbox{\vphantom{$\left<#2#3\right|$}}
\right. #3 \right> }}

\title{Absolute $^3$He polarimetry for a double-chambered cell using transmission of thermal neutrons}  

\author{A.~Watanabe$^1$}
\email{atomu@phys.titech.ac.jp}

\author{K.~Sekiguchi$^{1,2}$}
\author{T.~Ino$^3$}
\author{M.~Inoue$^2$}
\author{S.~Nakai$^{2,4}$}
\author{Y.~Otake$^5$}
\author{A.~Taketani$^5$}
\author{Y.~Wakabayashi$^5$}


\affiliation{$^1$Department of Physics, Tokyo Institute of Technology, Tokyo 152-8551, Japan} 
\affiliation{$^2$Department of Physics, Tohoku University, Sendai 980-8578, Japan}
\affiliation{$^3$High Energy Accelerator Research Organization (KEK), Tsukuba 305-0801, Japan}
\affiliation{$^4$Graduate Program on Physics for the Universe (GP-PU), Tohoku University, Sendai 980-8578, Japan}
\affiliation{$^5$RIKEN Center for Advanced Photonics, RIKEN, Wako 351-0198, Japan}

\date{\today}% It is always \today, today,

\begin{abstract}
We present an absolute $^3$He polarimetry method based on thermal neutron transmission for a double-chambered cell.
This method utilizes the fact that a $^3$He nucleus has a large absorption cross section and a spin dependence for thermal neutrons.
The cell had a pumping chamber and a target chamber.
Polarized $^3$He gas was produced in the pumping chamber by SEOP and then diffused into the target chamber.
The $^3$He polarization in the target chamber was determined by comparing the neutron transmissions with the polarized and unpolarized targets.
The measurement was performed at the RIKEN Accelerator-Driven Compact Neutron Source.
The $^3$He polarization in the target chamber was determined with a statistical error of 1.8\% and systematic uncertainty of 0.6\%.
This method can be used to obtain high-precision data of spin observables in few-nucleon scattering for the investigation of nuclear forces.
\end{abstract}

\maketitle

% Main text
\section{Introduction}
\label{sec:intro}
Nuclear-polarized $^3$He is widely used in several fields in scientific research.
The nuclear spin of $^3$He is predominantly the neutron spin~\cite{PhysRevC.42.2310}.
Hence, a polarized $^3$He target is used in electron scattering for measurements of spin-dependent observables involving neutrons~\cite{PhysRevC.91.055205}.
Polarized $^3$He is also utilized as a neutron spin filter~\cite{COULTER1990463}, and it has been used for high-accuracy magnetometry~\cite{PhysRevLett.124.223001}.
Furthermore, polarized $^3$He target has proven to be a useful target in the study of nuclear interactions from few-nucleon scattering~\cite{PhysRevC.103.044001}.

An accurate measurement of absolute target polarization is essential in the study of nuclear forces by few-nucleon scattering.
Recently, rigorous numerical calculations for four-nucleon (4$N$) scattering based on realistic nuclear potentials have become possible, and the demand for high-precision scattering data for comparison with these calculations is increasing.
The existing data of few-nucleon scattering typically have an accuracy of approximately 5\%.
Therefore, in the case of spin observables, an uncertainty of less than a few percent is required for the target polarization.

Recently, proton-$^3$He ($p$-$^3$He) elastic scatterings at 65, 70 and 100~MeV have been measured with the aim of ascertaining the three-nucleon force (3NF) effects at the Research Center for Nuclear Physics (RCNP), Osaka University, and the Cyclotron Radioisotope Center (CYRIC), Tohoku University~\cite{PhysRevC.103.044001, Atomu_PhD, Nakai_dthesis}.
The results indicate that the $p$-$^3$He scattering at intermediate energies is an excellent tool for the study of 3NFs that is not accessible by three-nucleon scattering.
A polarized $^3$He target system developed for scattering experiments has been applied to measure the spin observables.
The target container is a cell consisting of two chambers: a pumping chamber and a target chamber. 
%The pumping section polarizes $^3$He gas, and it diffuses into the target section through which a beam passes.
$^3$He gas is polarized in the pumping section and it diffuses into the target section through which a beam passes.
Consequently, this results in a lower polarization in the target section than that in the pumping section.
%This is so-called a polarization gradient, that is a known problem in a polarized $^3$He target system employing a double-chambered cell~\cite{PhysRevC.84.065201}.
This polarization gradient is a known problem in a polarized $^3$He target system employing a double-chambered cell~\cite{PhysRevC.84.065201}.

Nuclear magnetic resonance (NMR) and electron paramagnetic resonance (EPR)~\cite{PhysRevA.58.3004} are used to measure the $^3$He polarization.
The NMR signal, that produces relative polarization at the target section, generally needs to be calibrated using other methods.
In the EPR polarimetry, the $^3$He polarization is obtained by measuring the EPR frequency shift of alkali metal atoms in the presence of polarized $^3$He gas.
Although the EPR polarimetry yields the absolute values of the $^3$He polarization, it can be performed only if a mixture of $^3$He gas and alkali-metal vapor is present in the pumping section.
Therefore, it is necessary to determine the target polarization using a new $^3$He polarimetry technique to achieve high precision of polarization observables. 

In a neutron spin filter, neutron transmission depends on the $^3$He polarization because absorption occurs only if the neutron and $^3$He spins are antiparallel.
Hence, we measured the absolute $^3$He polarization in the target chamber by utilizing the transmission of thermal neutrons.
The target polarization was accurately measured by comparing the neutron transmissions with the polarized and unpolarized $^3$He gases.
Based on the result of the neutron transmission measurement, the absolute NMR calibration was performed for the $p$-$^3$He scattering experiments~\cite{PhysRevC.103.044001, Atomu_PhD, Nakai_dthesis}.

In this paper, we propose $^3$He polarimetry for a double-chambered cell by thermal neutron transmission using a neutron source at RIKEN.
In Section~2, we describe the principle of $^3$He polarimetry using neutron transmission.
The polarized $^3$He target system developed for scattering experiments is presented in Section~3.
Section~4 presents the experimental procedure, and Section~5 presents the experimental results and data analysis.
Finally, we summarize and conclude this work in Section~6.

\section{$^3$He polarimetry using neutron transmission}
The absorption cross section of thermal neutrons is quite large in $^3$He nuclei and is spin dependent~\cite{COULTER198890}.
Therefore, neutron transmission for polarized $^3$He nuclei $T_n$ is expressed as
\begin{eqnarray}
	T_n &=& e^{-\mathcal{O}} \cosh (P_{\rm He} \mathcal{O}),
	\label{n_trans} \\
	\mathcal{O} &=& \sigma_{\rm abs} n_{\rm He} d,
	\label{opacity}
\end{eqnarray}
where $\mathcal{O}$ is the opacity, $\sigma_{\rm abs}$ is the spin-averaged neutron absorption cross section of $^3$He, $n_{\rm He}$ is the $^3$He number density, $d$ is the thickness of the $^3$He gas, and $P_{\rm He}$ is the $^3$He polarization, respectively.
We ignore the elastic scattering cross section (3~barn \cite{SovJNuclPhys.25.607}), which is sufficiently small compared to $\sigma_{\rm abs}$ at the energies of interest here.
$\sigma_{\rm abs}$ was accurately measured by Keith $et~al.$ at low energies (0.1--400~eV) \cite{PhysRevC.69.034005}:
\begin{equation}
	\sigma_{\rm abs} = (849.77 \pm 0.14_{\rm (sta)} \pm 1.02_{\rm (sys)}) E_n^{-1/2}
					   -(1.25 \pm 0.00_{\rm (sta)} {}^{+0.01}_{-0.05 \rm (sys)})~[{\rm barn}],
	\label{cs_abs}
\end{equation}
where $E_n$ is the neutron energy (eV).
The first uncertainty is statistical and the second is systematic. 
Therefore, the $^3$He polarization is calculated as
\begin{equation}
	P_{\rm He} = -\frac{1}{\ln T_{n,0}} \cosh^{-1} \left( \frac{T_n}{T_{n,0}} \right),
	\label{3he_pol_trans}
\end{equation}
where $T_{n,0}$ is the neutron transmission through the unpolarized $^3$He gas.
%Thus, the $^3$He polarization in the target chamber can be obtained from the ratio of $T_n$ to $T_{n,0}$.
Thus, the $^3$He polarization in the target chamber can be obtained from the ratio of $T_n$ to $T_{n,0}$, and the value of $T_{n,0}$.

\section{Polarized $^3$He target}
\label{sec:pol3He}
\begin{figure}[tbp]
 \centering
  \begin{tabular}{c}
  
	\begin{minipage}{0.4\hsize}
	 \centering
	  	\includegraphics[clip, width=7.0cm]{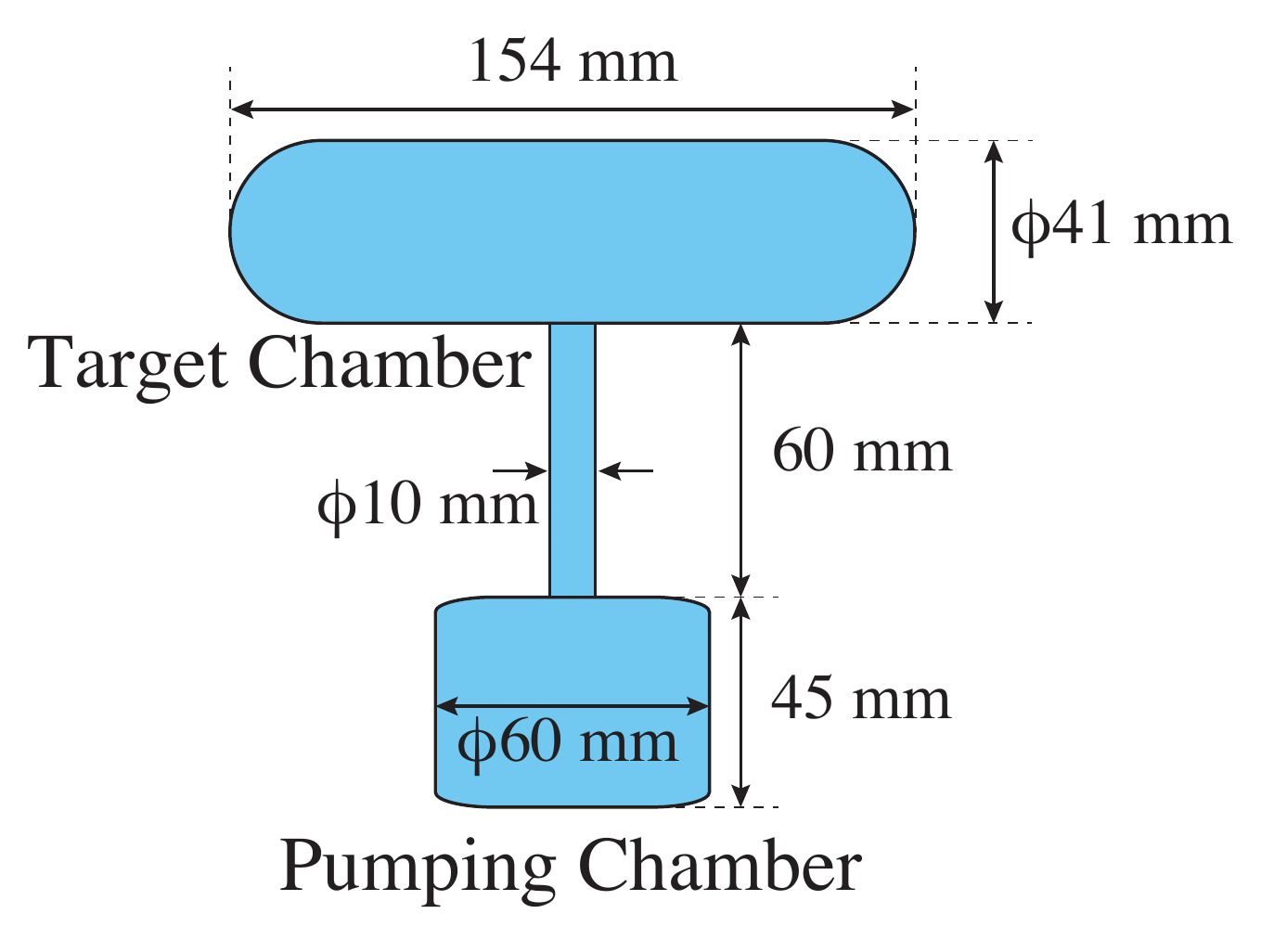}
	\end{minipage}
%	\hspace{-0.5cm}
	\begin{minipage}{0.4\hsize}
	 \centering
	 	\includegraphics[clip,width=4.5cm]{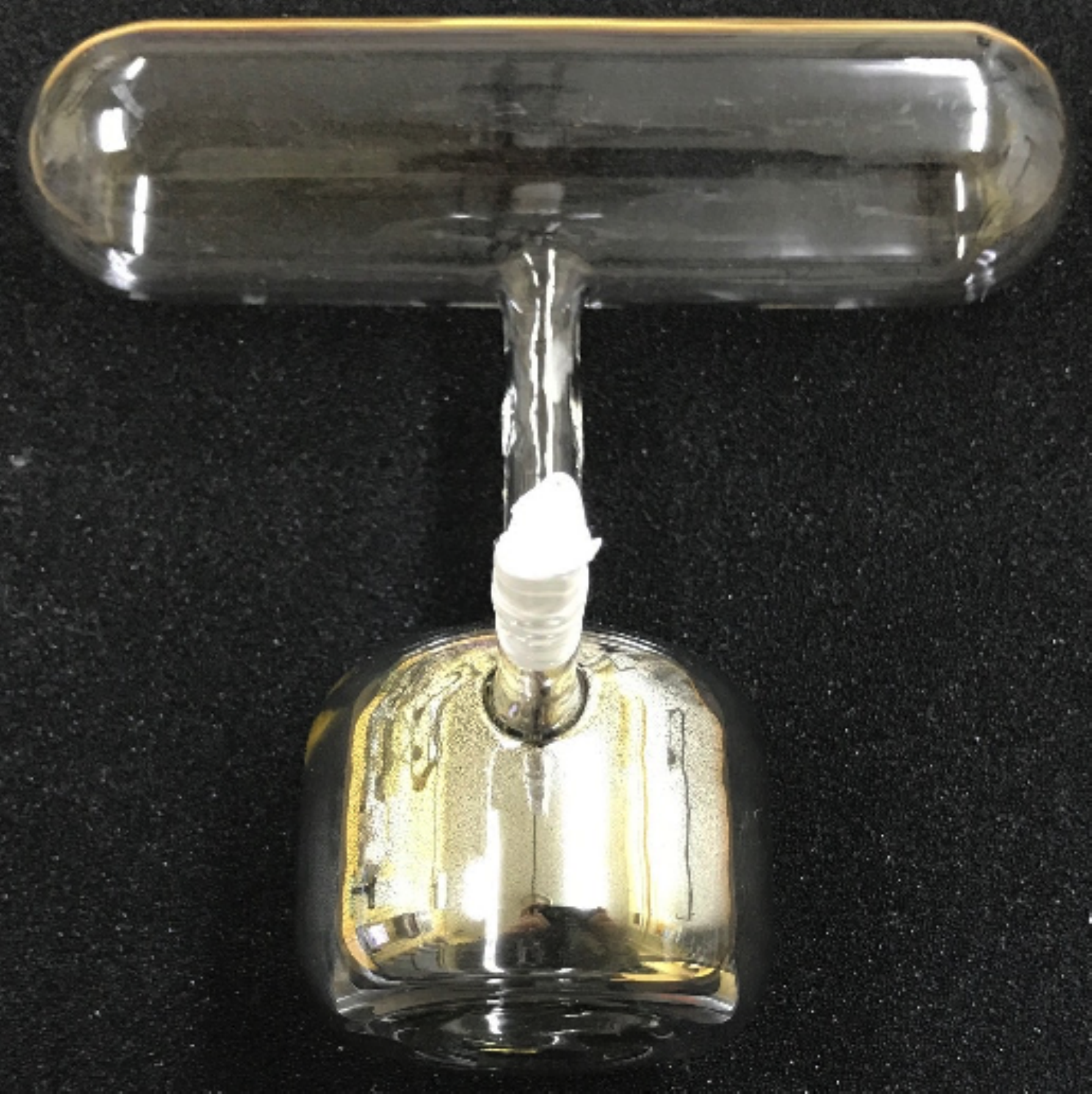}
	\end{minipage}
	
  \end{tabular}
 \caption{Dimensions and photograph of the target cell.}
 \label{cell_dim}
\end{figure}

Polarized $^3$He nuclei were produced using the spin-exchange optical pumping (SEOP) method~\cite{PhysRevLett.5.373,PhysRevLett.91.123003}.
In SEOP, a vapor of alkali metal atoms is optically pumped, which transfers its polarization to $^3$He nuclei by spin-exchange interactions.
The dimensions of the target cell are shown in Fig.~\ref{cell_dim}.
The target cell was made of boron-free aluminosilicate glass (GE180), which consisted of a target chamber and a pumping chamber connected by a transfer tube.
$^3$He gas was polarized in the pumping chamber by SEOP and then diffused into the target chamber.
This design prevented the depolarization of alkali-metal atoms due to the charged particle beam~\cite{COULTER198929} and the undesirable energy loss of scattered particles through a material that was used to heat the pumping chamber.
The target cell contained $^3\rm He$ gas with a pressure of 3~atm at room temperature, a small amount of $\rm N_2$ gas (0.1~atm), and a mixture of Rb and K alkali metals.
The ratio of the alkali metal number densities $\rm{[K]/[Rb]}$ was 3.4 at 500~K obtained by the white light absorption~\cite{RevModPhys.89.045004}.
The target chamber had a length and diameter of 15.4~cm and 4.1~cm, respectively, whereas the pumping chamber had corresponding dimensions of 4.5~cm and 6.0~cm, respectively.
The entrance and exit windows had a thickness of 0.4~mm, and the thickness of the side surfaces where scattered protons passed was approximately 1~mm.

\begin{figure}[tbp]
\begin{center}
	\includegraphics[width=10cm]{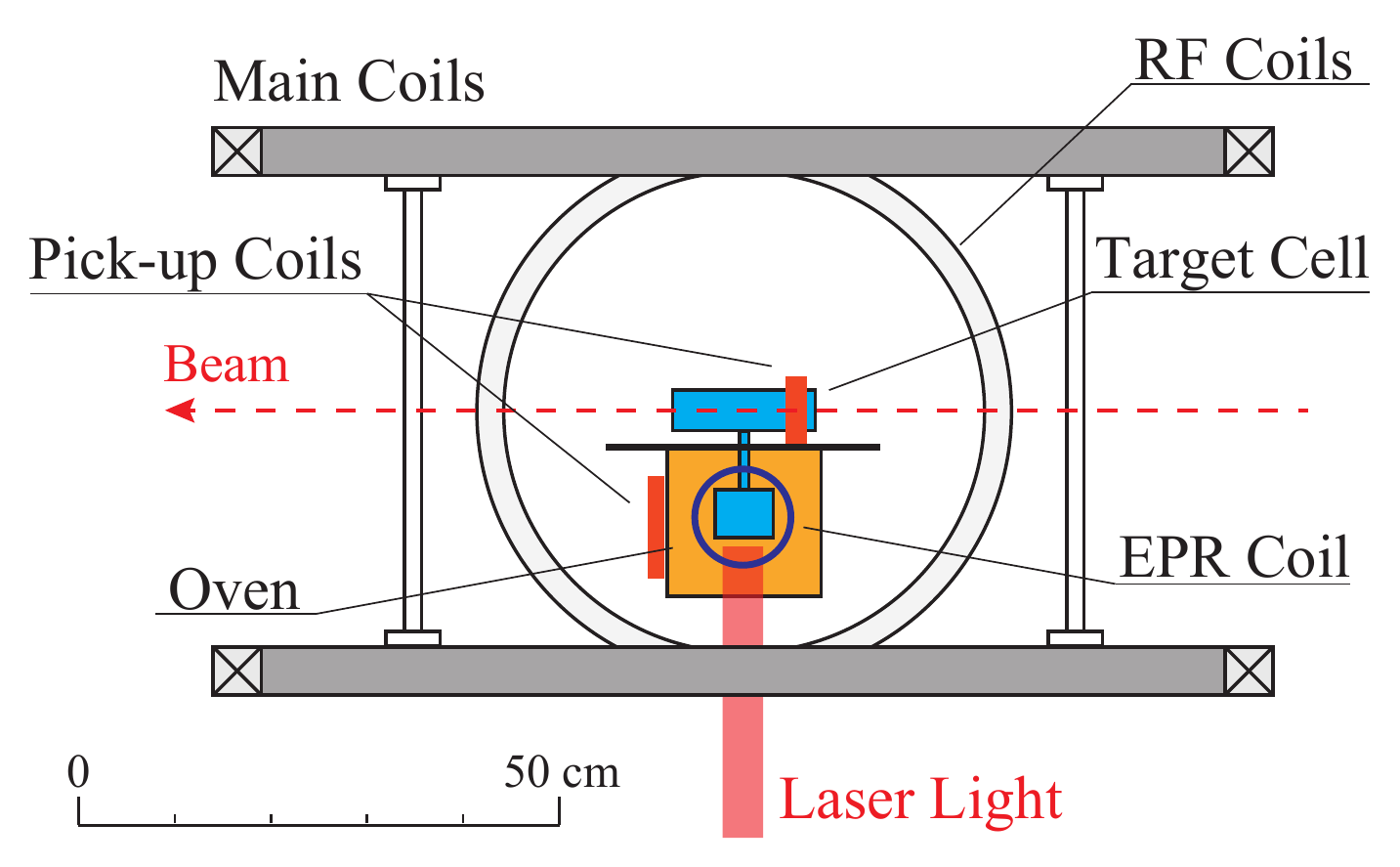}
	\caption{Schematic view of the polarized $^3$He target system.}
	\label{fig:pol-sys}
\end{center}
\end{figure}

Figure~\ref{fig:pol-sys} shows a schematic view of the polarized $^3$He target system.
A 1.2~mT magnetic field, produced by a pair of Helmholtz coils having a diameter of 100~cm (main coils), defined the direction of the $^3$He nuclear polarization.
The target cell was placed at the center of the main coils.
The pumping chamber was heated to 500~K using a hot air blower to achieve a sufficient amount of alkali metal vapor density.
A diode laser bar with an output power of 65~W \cite{Coherent-DILAS}  was used for SEOP.
%Diode laser was used for SEOP with the maximum output power of 65~W.
Rb atoms were polarized by optical pumping with circularly polarized light at 794.7~nm in the pumping chamber.
The linewidth of the laser was narrowed to about 0.2 nm (FWHM) by an external cavity with a volume Bragg grating~\cite{JApplPhys.116.014903}.
The temperature of the target cell was measured with thermocouples attached to the surfaces of the target and pumping chamber, respectively.

%RF coils and pick-up coils were used for NMR polarimetry, and an EPR coil was used for EPR polarimetry.
The $^3$He polarization was measured using both the adiabatic fast passage NMR (AFP-NMR) and the EPR methods.
The AFP-NMR was used to monitor the relative $^3$He polarization, which included a pair of 45~cm diameter Helmholtz coils that generated RF field (RF coils) and two pick-up coils.
The RF field was normal to the direction of the  static magnetic field.
The two pick-up coils, which were perpendicular to both the static magnetic field and the RF field, were used to detect the induced NMR signal.
One was placed so as to cover the target chamber, and the other was placed on the side surface of the oven for the pumping chamber.
The EPR method was used to roughly estimate the $^3$He polarization in the pumping chamber.
The EPR system consisted of a EPR coil, which was placed near the pumping chamber in the oven, to provide an RF field and a photodiode detector to measure the D2 fluorescent light from Rb (780~nm) with a band-pass filter.
The output of the photodiode detector was connected to a PC-based oscilloscope which had a build-in function generator (FG). 
The FG output of the oscilloscope was used as a source of the RF filed.
The RF frequency was modulated with a triangular function in the build-in FG, and the feedback by phase-sensitive detection of the input signal was applied to match the EPR frequency.
See Ref.~\cite{Ino_2019} for details of the EPR system.
The typical $^3$He polarization in the pumping chamber was 40\%, and the spin relaxation time in the cell was 35~h.
The spin-up (spin-relaxation) time constant in the cell was 9.2~h (35~h) in our conditions.

\section{Experimental procedure}
\begin{figure*}[tbp]
	\centering
	\includegraphics[width=16.5cm]{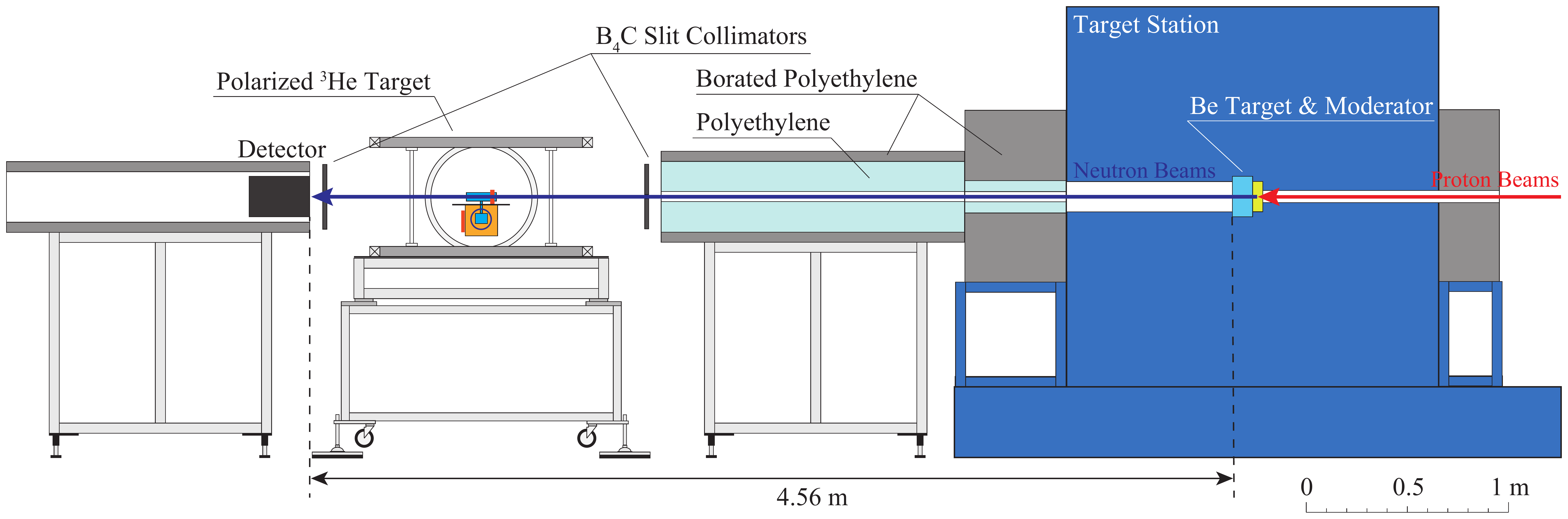}
	\caption{Experimental setup of the neutron transmission measurement.}
	\label{setup_rans}
\end{figure*}

The measurement of the absolute $^3$He polarization in the target chamber was performed at the RIKEN Accelerator-driven compact Neutron Source (RANS).
Figure~\ref{setup_rans} shows the experimental setup of the neutron transmission measurement.
The RANS was built at RIKEN for neutron scattering measurements and other applications \cite{IKEDA201661}.
The RANS includes a proton linear accelerator and a target station which consists of the Be target system for neutron production, carbon for neutron reflector, borated polyethylene for neutron shield, and lead for gamma-ray shield. 
Protons were pulsed and extracted from an ion source and accelerated by a linac up to 7~MeV.
Proton beams were injected into the Be target with a thickness of 0.3~mm placed inside the target station.
Neutrons were produced by the charge-exchange reaction Be($p, n$), and were moderated in a polyethylene moderator with a thickness of 40~mm.
Thus, the thermal neutrons were extracted from the target station.
The energy distribution of the neutrons had peaks at approximately 1.5~MeV and 50~meV \cite{IKEDA201661}.
The neutron beams from the target station were transported to the polarized $^3$He target through a neutron collimator composed of borated polyethylene, because boron has a large absorption cross section for neutrons.
The polarized $^3$He target was installed downstream of the neutron collimator.
We also installed B$_4$C slit collimators at the front and back of the polarized $^3$He target so that the neutron beams passed only through the center of the target chamber.
The B$_4$C slit collimator consists of four sintered B$_4$C plates with a thickness of 5~mm mounted on a finely movable stage.
The collimation size of these B$_4$C slit collimators was set to $10 \times 10~{\rm mm^2}$ in this experiment.
The neutron detector was placed downstream of the target and was covered by a B$_4$C sheet, except for the neutron beam, to reduce background neutrons.
The distance from the moderator surface to the detector was 4.56~m.
In the experiment, the average proton beam current, repetition rate, and pulse width of the proton beams were 20~$\rm \mu A$, 125~Hz, and 20~$\rm \mu s$, respectively.

The neutron detector consisted of a ZnS(Ag)/$^6$LiF scintillator optically coupled with a position-sensitive photo multiplier tube (RPMT).
A ZnS(Ag)/$^6$LiF scintillator had a large light yield for thermal neutrons.
This RPMT had a 12-stage mesh dynode and an anode wire structure crossing in the $x$-axis and $y$-axis directions, respectively.
Output currents from each anode wire were divided into two signals by resistors.
The positions of the detected neutrons were obtained from the ratio of these currents \cite{PCCP.7.1836}.
Specifications of the neutron detector are shown in Table~\ref{spe_n_det}.
The neutron energy was determined by the time-of-flight (ToF) between a signal from the proton linac and the event trigger of the detector.
Here, we assumed that thermal neutrons can be treated non-relativistically.
Thus, the neutron energy is expressed as
\begin{equation}
	E_n = \frac{1}{2} m_n \left( \frac{L}{t} \right)^2,
	\label{En_nonrelative}
\end{equation}
where $m_n$ is the neutron mass, $L$ is the distance from the moderator surface to the neutron detector, and $t$ is the ToF.

\begin{table}[tbp]
\caption{Specifications of the neutron detector.}
\centering
\begin{tabular}{c|c} \hline \hline
	Scintillator & ZnS(Ag)/$^6$LiF \\
	Scintillator thickness & $0.25~{\rm mm}$ \\
	Effective area & $\phi = 90~{\rm mm}$ \\
	Efficiency & $30\%$ (for cold neutron) \\
	Spatial resolution & $0.8~{\rm mm}$ (FWHM) \\ \hline \hline
\end{tabular}
\label{spe_n_det}
\end{table}%

We performed the neutron transmission measurements under several target conditions:
(A) no target (direct beam), 
(B) a blank cell with nearly the same dimensions as those of the target cell, 
(C) an unpolarized target cell under the operating conditions (the temperature in the cell was 500~K with the pumping laser irradiation), 
(D) a polarized target cell under the operating conditions, and 
(E) background measurements.
Measurements (A) and (B) were used to estimate the neutron transmission through the glass windows of the blank cell.
The neutron transmission through the glass windows of the $^3$He cell was obtained by correcting that of the blank cell from the difference in the window thicknesses of the blank cell and the $^3$He cell.
The glass-window thicknesses were measured with an ultrasonic thickness gauge.
Measurements (C) and (D) were used to determine the $^3$He polarization using Eq.~\eqref{3he_pol_trans}.
During the measurement (C), we used NMR to ensure that the $^3$He gas remained unpolarized.
The transmission measurement (D) was conducted continuously for three days, starting at the beginning of the optical pumping with a build-up time of 10~h through the saturation of the polarization. 
For the measurement (E), we placed a B$_4$C sheet with a thickness of 13~mm between the target cell and the neutron detector to shield from the neutron beams.

The neutron transmission passing through the $^3$He gas in a target cell can be rewritten as
\begin{equation}
	T_n = \frac{N_{\rm He} - N_{\rm BG}}{N_{\rm Blank} - N_{\rm BG}},
	\label{n_trans_2}
\end{equation}
where $N$ is the number of detected neutrons.
The subscripts He, Blank, and BG denote the measurements for the $^3$He cell, blank cell, and background, respectively.
Note that $N_{\rm He}$, $N_{\rm Blank}$, and $N_{\rm BG}$ were normalized by the proton beam intensity.

\section{Experimental results and data analysis}
\begin{figure*}[tbp]
\centering
	\includegraphics[clip,width=14cm]{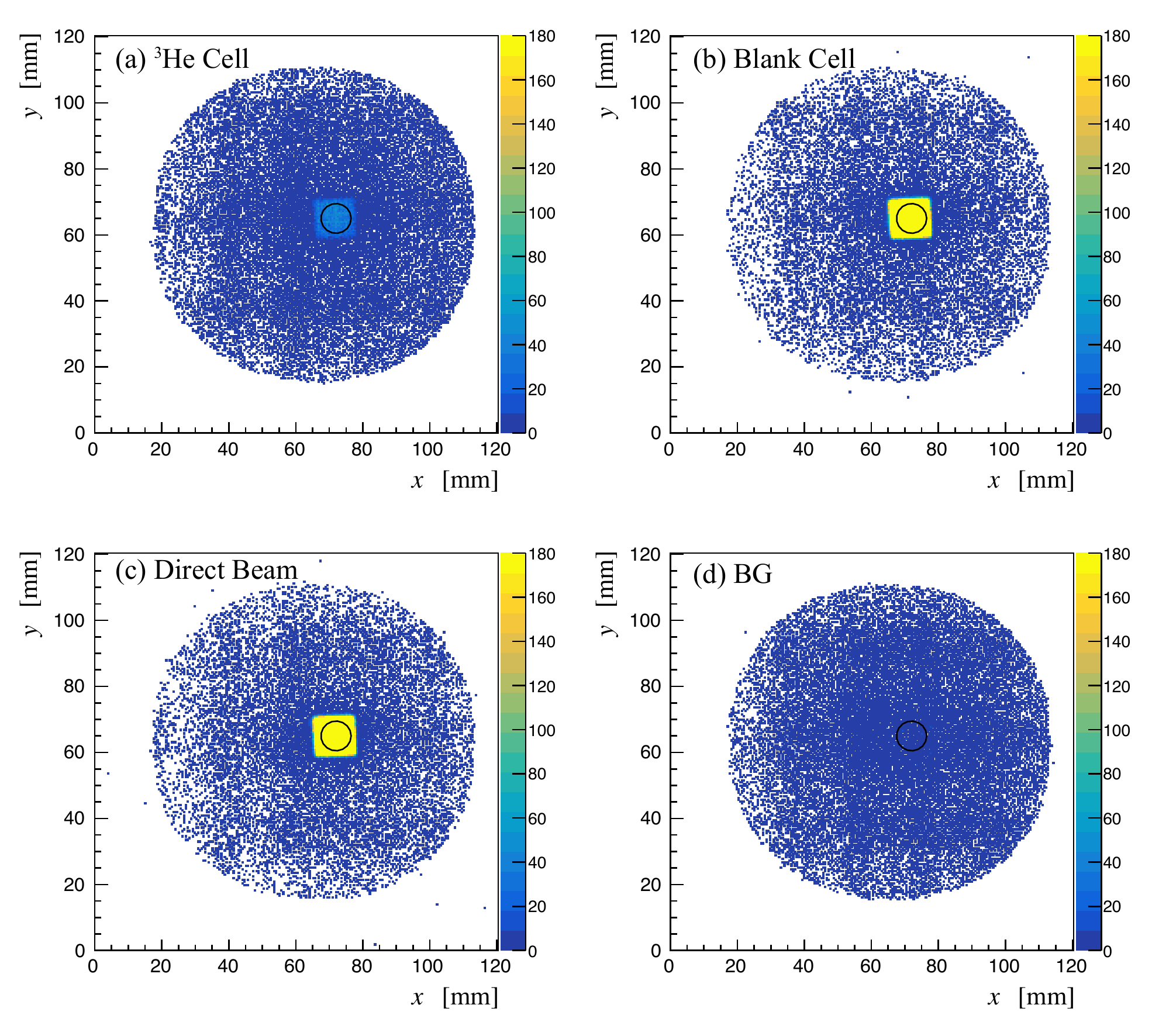}
\caption{Typical two-dimensional images of the detected neutron positions of the transmission measurement for the (a) $^3$He target cell, (b) blank cell, (c) direct beam measurement, and (d) background (BG) measurement.
		 Neutrons within the solid circles, with diameters of 9~mm, were used for the analysis.
		 The ToF gate was not applied here.}
\label{pos2d}
\end{figure*}

Figure~\ref{pos2d} shows typical two-dimensional images of the detected neutron positions for the measurements under each of the following conditions: $^3$He target cell, a blank cell, no target, and background.
The neutrons passing through the B$_4$C slit collimator with an area of $10 \times 10~{\rm mm^2}$ are clearly visualized in panels (a)--(c) of Fig.~\ref{pos2d}.
We selected neutrons within a $\phi 9~{\rm mm}$ circle in the center of this area to detect the neutrons that passed through the center of the target chamber.

\begin{figure}[tbp]
	\centering
	\includegraphics[clip,width=10cm]{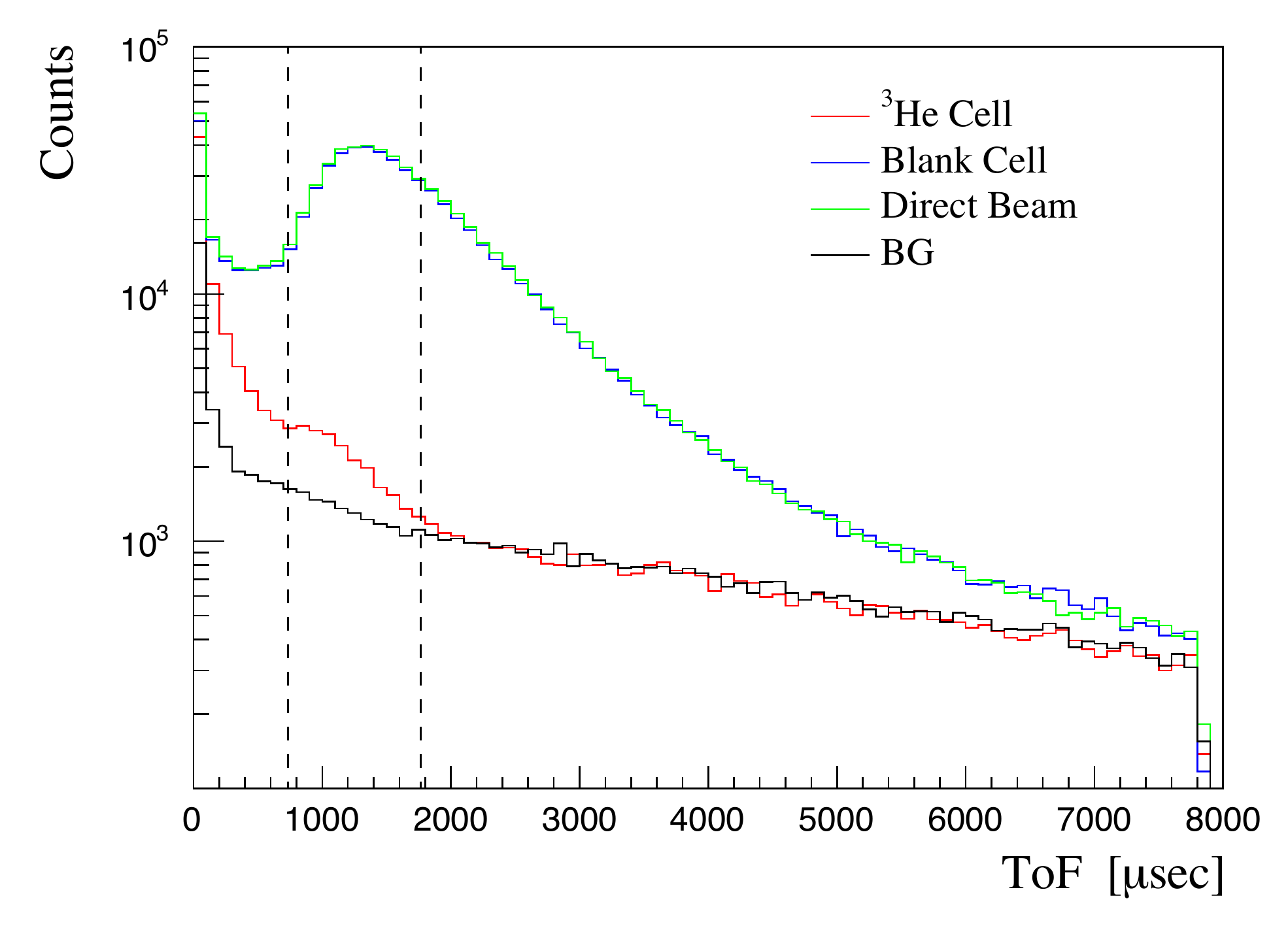}
 	\caption{Typical time-of-flight (ToF) spectra for the transmission measurements with the $^3$He cell (red), blank cell (blue), direct beam measurement (green), and background measurement (black).
		 The vertical dashed lines represent the gates that eliminate slow and fast neutrons.}
	\label{tof}
\end{figure}

\begin{figure}[tbp]
	\centering
	\includegraphics[clip,width=10cm]{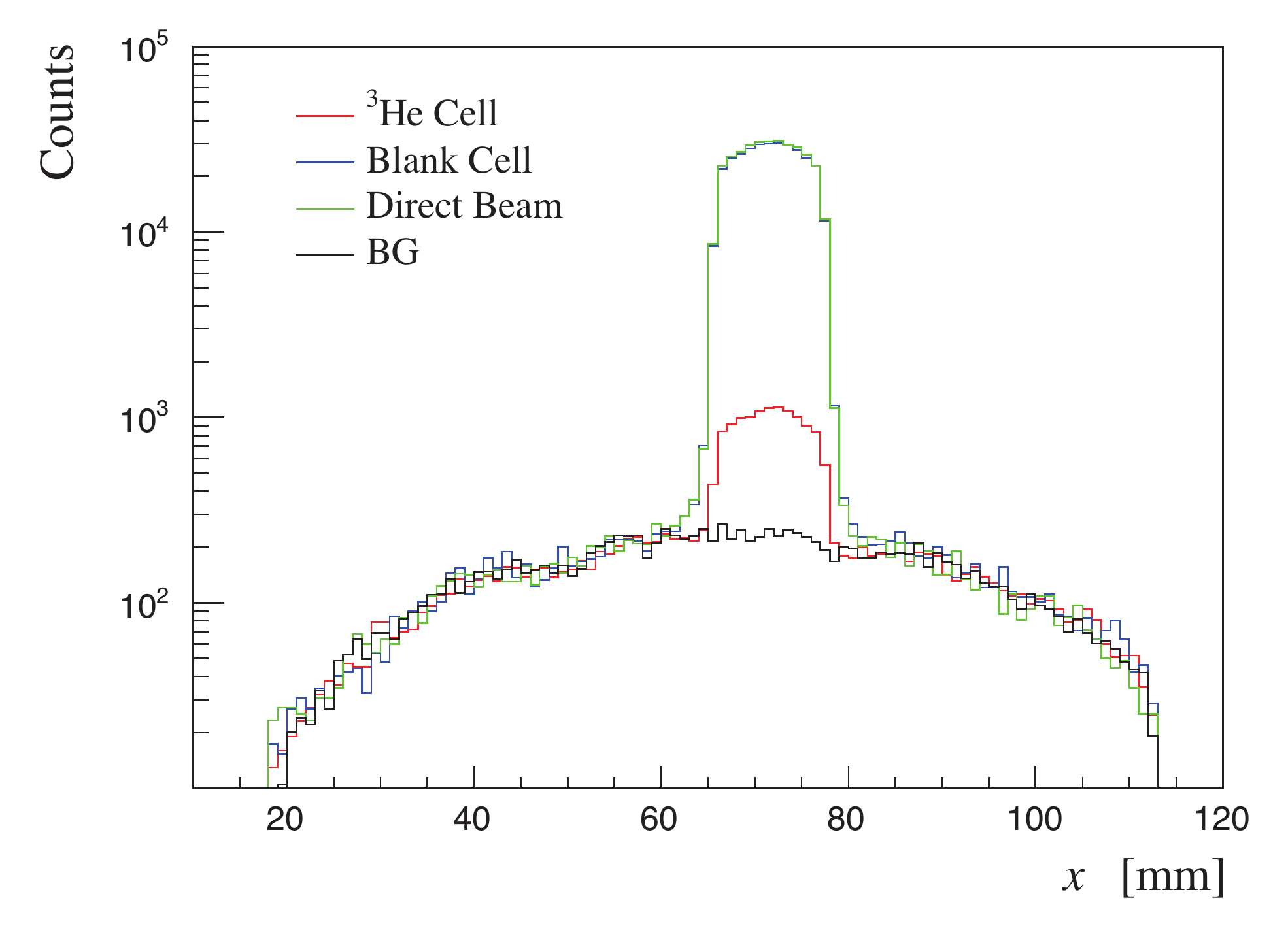}
 	\caption{Projections of the detected neutron positions on the $x$-axis for the transmission measurements with the $^3$He cell (red), blank cell (blue), direct beam measurement (green), and background measurement (black).
		 The neutrons were selected in the range of $E_n = 35~{\rm meV}$ (${\rm ToF}=1762~{\rm \mu s}$) to $200~{\rm meV}$ (${\rm ToF}=737~{\rm \mu s}$).
		 Each count of detected neutrons was normalized by the proton beam intensity.}
	\label{xpos}
\end{figure}

Figure~\ref{tof} shows typical ToF spectra of detected neutrons in the entire effective area of the detector.
The slow neutrons were completely absorbed by the $^3$He cell, and fast neutrons penetrated the B$_4$C sheet.
Thus, we selected neutrons in the range of $E_n = 35~{\rm meV}$ (${\rm ToF}=1762~{\rm \mu s}$) to $200~{\rm meV}$ (${\rm ToF}=737~{\rm \mu s}$) to eliminate these neutrons.
The projections of the neutron positions on the $x$-axis of Fig.~\ref{pos2d} are shown in Fig.~\ref{xpos}.
The numbers of the detected neutrons were normalized by the proton intensities.
The backgrounds were removed by subtracting the events obtained from the background measurement.

\begin{figure}[tbp]
	\centering
	\includegraphics[width=10cm]{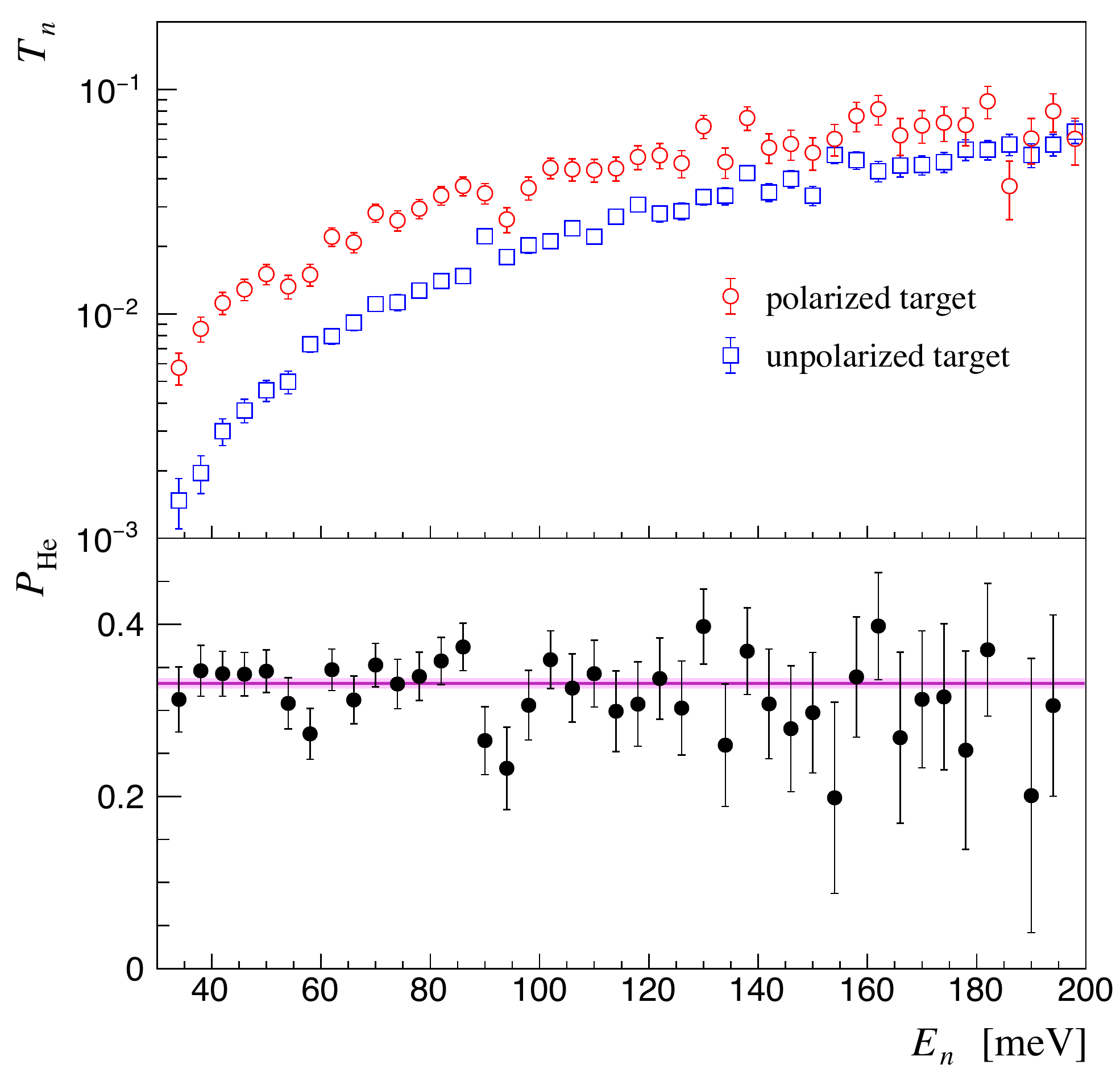}
	\caption{Energy dependence of neutron transmission for the polarized (red circles) and unpolarized (blue squares) target cells (top panel), and the $^3$He polarization (bottom panel).
	Statistical errors are shown for each panel.
	The magenta solid line shows the weighted mean, and the light band shows the statistical error.}
	\label{trans_3he}
\end{figure}

The top panel of Fig.~\ref{trans_3he} shows the neutron transmission results obtained from Eq.~\eqref{n_trans_2} for the unpolarized and polarized cells at neutron energies of 35--200~meV.
The neutron transmission through the polarized $^3$He target clearly increased compared to that of the unpolarized target.
%The resulting $^3$He polarization in the target chamber calculated using Eq.~\eqref{3he_pol_trans} is also shown in the bottom panel of Fig.~\ref{trans_3he}.
The energy-independent $^3$He polarization in the target chamber, calculated using Eq.~\eqref{3he_pol_trans}, is also shown in the bottom panel of Fig.~\ref{trans_3he}.
By taking the weighted average value in the energy range of 35--200~meV, the absolute $^3$He polarization of the target chamber was obtained as
\begin{equation}
	P_{\rm He} = 0.331 \pm 0.006_{\rm (sta)} \pm 0.002_{\rm (sys)},
	\label{Ptc_ishi}
\end{equation}
with a statistical error of 1.8\%.
The systematic error was estimated to be 0.6\%.
The main contribution to the systematic error is the uncertainty of the proton beam intensity.
We evaluated this uncertainty from the deviation of the average pulse current for each measurement because the measurements were performed under the same beam conditions throughout the experiment.
The other contribution was due to the difference in the thickness of the glass windows of the blank cell and the $^3$He cell.
This difference was applied to obtain true neutron transmissions for $^3$He.
The neutron transmission through the blank cell was determined to be $0.985 \pm 0.003$ from the measurements (A) and (B).
As a result, the transmission through the glass windows of the $^3$He cell was obtained as $0.981 \pm 0.004$ after correction for the glass thickness measurement.
We estimated that the contribution from these uncertainties to the $^3$He polarization was 0.1\%; thus, the contribution was negligibly small.

We also performed the measurement of $^3$He number densities in the same experimental setup as the $^3$He polarization measurement using neutron transmission.
The average $^3$He number density $n_0$ of the target cell was obtained as,
\begin{equation}
	n_0 = (8.02 \pm 0.05_{\rm (sta)} \pm 0.19_{\rm (sys)}) \times 10^{19}~~[{\rm cm^{-3}}].
	\label{dens_ishi_0}
\end{equation}
by the neutron transmission measurement for an unpolarized target cell at room temperature.
The statistical error was 0.6\%.
The systematic uncertainty was estimated to be 2.4\%, which was the quadratic sum of the uncertainties in the ToF, the inner length of the cell and the proton beam intensity.
The $^3$He number density in the target chamber under the operating condition $n_{\rm tc}$ was also measured in the same manner,
\begin{equation}
	n_{\rm tc} = (9.20 \pm 0.04_{\rm (sta)} \pm 0.22_{\rm (sys)}) \times 10^{19}~~[{\rm cm^{-3}}],
	\label{dens_ishi_operate}
\end{equation}
with the statistical error of 0.4\%.
The systematic uncertainty was estimated to be 2.4\% same as the average $^3$He number density.

\section{Conclusions}
We report on the absolute $^3$He polarimetry for a double-chambered cell by thermal neutron transmission.
This polarimetry method was based on the fact that a $^3$He nucleus has a larger absorption cross section for thermal neutrons with strong spin dependence.
The measurement was carried out at RANS with a polarized $^3$He target system developed for $p$-$^3$He scattering experiments.
The $^3$He polarization in the target chamber was determined by comparing the neutron transmissions with the polarized and unpolarized $^3$He targets.
The statistical error was 1.8\% and the systematic uncertainty was estimated to be 0.6\%.
Taking the NMR signals before and after the neutron transmission measurements, the calibration coefficients, the ratios of the NMR signal height to the absolute $^3$He polarization, were determined.
This result was used to evaluate the target $^3$He polarization and to obtain high-precision data of the spin observables described in Ref.~\cite{PhysRevC.103.044001, Atomu_PhD}.
The findings of this study was also used to measure the spin correlation coefficient $C_{y,y}$ at 100~MeV~\cite{Nakai_dthesis}, which was obtained using a polarized proton beam and a polarized $^3$He target.

In a double-chambered cell using SEOP, the temperature distribution inside the cell is known to be highly complex because of additional heating by laser irradiation, which is a major source of uncertainty in the EPR polarimetry.
Normand $et~al.$ measured the $^3$He number density corresponding to the temperature inside a cell by using neutron transmission \cite{Normand_2016}.
They observed a $\rm 135^\circ C$ temperature rise in a 1~atm $^3$He cell under illumination by a 200~W pumping laser.
We estimated $^3$He number densities in each chamber by performing a series of neutron transmission measurements in conjunction with the average $^3$He number density of the target cell.
Similarly, the ratio of the temperatures inside the target and pumping chambers was deduced.
In our case, the temperature rise inside the cell was estimated to be about 10$^\circ \rm C$ compared to the temperature on the cell surface.
By using the obtained $^3$He number densities and temperatures in the two chambers, the absolute $^3$He polarization in the pumping chamber can be accurately measured by EPR.
Note that the absolute $^3$He polarization in the target chamber was obtained from neutron transmissions.
Further studies on the $^3$He gas diffusion inside the target cell will help enhance the quantitative understanding of the $^3$He polarization difference between two chambers.
Further research will help achieve $^3$He polarization with sufficient accuracy for the investigation of nuclear interactions from few-nucleon scattering without performing other neutron transmission measurements.

%\textcolor{red}{}

\section*{Acknowledgment}
We acknowledge the help and support of the RANS team for performing our experiments.
This work was financially supported in part by JSPS KAKENHI Grant Nos. JP25105502, JP16H02171, and JP18H05404.


%\bibliographystyle{cas-model2-names}
%\bibliographystyle{apsrev_mod}

%\bibliography{bibliography}

\begin{thebibliography}{22}
\expandafter\ifx\csname natexlab\endcsname\relax\def\natexlab#1{#1}\fi
\expandafter\ifx\csname bibnamefont\endcsname\relax
  \def\bibnamefont#1{#1}\fi
\expandafter\ifx\csname bibfnamefont\endcsname\relax
  \def\bibfnamefont#1{#1}\fi
\expandafter\ifx\csname citenamefont\endcsname\relax
  \def\citenamefont#1{#1}\fi
\expandafter\ifx\csname url\endcsname\relax
  \def\url#1{\texttt{#1}}\fi
\expandafter\ifx\csname urlprefix\endcsname\relax\def\urlprefix{URL }\fi
\providecommand{\bibinfo}[2]{#2}
\providecommand{\eprint}[2][]{\url{#2}}

\bibitem[{\citenamefont{Friar {\it et~al}.}(1990)\citenamefont{Friar, Gibson,
  Payne, Bernstein, and Chupp}}]{PhysRevC.42.2310}
\bibinfo{author}{\bibfnamefont{J.~L.} \bibnamefont{Friar}},
  \bibinfo{author}{\bibfnamefont{B.~F.} \bibnamefont{Gibson}},
  \bibinfo{author}{\bibfnamefont{G.~L.} \bibnamefont{Payne}},
  \bibinfo{author}{\bibfnamefont{A.~M.} \bibnamefont{Bernstein}},
  \bibnamefont{and} \bibinfo{author}{\bibfnamefont{T.~E.} \bibnamefont{Chupp}},
  \bibinfo{journal}{Phys. Rev. C} \textbf{\bibinfo{volume}{42}},
  \bibinfo{pages}{2310} (\bibinfo{year}{1990}).

\bibitem[{\citenamefont{Singh {\it et~al}.}(2015)\citenamefont{Singh, Dolph,
  Tobias, Averett, Kelleher, Mooney, Nelyubin, Wang, Zheng, and
  Cates}}]{PhysRevC.91.055205}
\bibinfo{author}{\bibfnamefont{J.~T.} \bibnamefont{Singh}},
  \bibinfo{author}{\bibfnamefont{P.~A.~M.} \bibnamefont{Dolph}},
  \bibinfo{author}{\bibfnamefont{W.~A.} \bibnamefont{Tobias}},
  \bibinfo{author}{\bibfnamefont{T.~D.} \bibnamefont{Averett}},
  \bibinfo{author}{\bibfnamefont{A.}~\bibnamefont{Kelleher}},
  \bibinfo{author}{\bibfnamefont{K.~E.} \bibnamefont{Mooney}},
  \bibinfo{author}{\bibfnamefont{V.~V.} \bibnamefont{Nelyubin}},
  \bibinfo{author}{\bibfnamefont{Y.}~\bibnamefont{Wang}},
  \bibinfo{author}{\bibfnamefont{Y.}~\bibnamefont{Zheng}}, \bibnamefont{and}
  \bibinfo{author}{\bibfnamefont{G.~D.} \bibnamefont{Cates}},
  \bibinfo{journal}{Phys. Rev. C} \textbf{\bibinfo{volume}{91}},
  \bibinfo{pages}{055205} (\bibinfo{year}{2015}).

\bibitem[{\citenamefont{Coulter {\it et~al}.}(1990)\citenamefont{Coulter,
  Chupp, McDonald, Bowman, Bowman, Szymanski, Yuan, Cates, Benton, and
  Earle}}]{COULTER1990463}
\bibinfo{author}{\bibfnamefont{K.}~\bibnamefont{Coulter}},
  \bibinfo{author}{\bibfnamefont{T.}~\bibnamefont{Chupp}},
  \bibinfo{author}{\bibfnamefont{A.}~\bibnamefont{McDonald}},
  \bibinfo{author}{\bibfnamefont{C.}~\bibnamefont{Bowman}},
  \bibinfo{author}{\bibfnamefont{J.}~\bibnamefont{Bowman}},
  \bibinfo{author}{\bibfnamefont{J.}~\bibnamefont{Szymanski}},
  \bibinfo{author}{\bibfnamefont{V.}~\bibnamefont{Yuan}},
  \bibinfo{author}{\bibfnamefont{G.}~\bibnamefont{Cates}},
  \bibinfo{author}{\bibfnamefont{D.}~\bibnamefont{Benton}}, \bibnamefont{and}
  \bibinfo{author}{\bibfnamefont{E.}~\bibnamefont{Earle}},
  \bibinfo{journal}{Nucl. Instr. and Meth. A} \textbf{\bibinfo{volume}{288}},
  \bibinfo{pages}{463} (\bibinfo{year}{1990}).

\bibitem[{\citenamefont{Farooq {\it et~al}.}(2020)\citenamefont{Farooq, Chupp,
  Grange, Tewsley-Booth, Flay, Kawall, Sachdeva, and
  Winter}}]{PhysRevLett.124.223001}
\bibinfo{author}{\bibfnamefont{M.}~\bibnamefont{Farooq}},
  \bibinfo{author}{\bibfnamefont{T.}~\bibnamefont{Chupp}},
  \bibinfo{author}{\bibfnamefont{J.}~\bibnamefont{Grange}},
  \bibinfo{author}{\bibfnamefont{A.}~\bibnamefont{Tewsley-Booth}},
  \bibinfo{author}{\bibfnamefont{D.}~\bibnamefont{Flay}},
  \bibinfo{author}{\bibfnamefont{D.}~\bibnamefont{Kawall}},
  \bibinfo{author}{\bibfnamefont{N.}~\bibnamefont{Sachdeva}}, \bibnamefont{and}
  \bibinfo{author}{\bibfnamefont{P.}~\bibnamefont{Winter}},
  \bibinfo{journal}{Phys. Rev. Lett.} \textbf{\bibinfo{volume}{124}},
  \bibinfo{pages}{223001} (\bibinfo{year}{2020}).

\bibitem[{\citenamefont{Watanabe {\it et~al}.}(2021)\citenamefont{Watanabe,
  Nakai, Wada, Sekiguchi, Deltuva, Akieda, Etoh, Inoue, Inoue, Kawahara {\it
  et~al}.}}]{PhysRevC.103.044001}
\bibinfo{author}{\bibfnamefont{A.}~\bibnamefont{Watanabe}},
  \bibinfo{author}{\bibfnamefont{S.}~\bibnamefont{Nakai}},
  \bibinfo{author}{\bibfnamefont{Y.}~\bibnamefont{Wada}},
  \bibinfo{author}{\bibfnamefont{K.}~\bibnamefont{Sekiguchi}},
  \bibinfo{author}{\bibfnamefont{A.}~\bibnamefont{Deltuva}},
  \bibinfo{author}{\bibfnamefont{T.}~\bibnamefont{Akieda}},
  \bibinfo{author}{\bibfnamefont{D.}~\bibnamefont{Etoh}},
  \bibinfo{author}{\bibfnamefont{M.}~\bibnamefont{Inoue}},
  \bibinfo{author}{\bibfnamefont{Y.}~\bibnamefont{Inoue}},
  \bibinfo{author}{\bibfnamefont{K.}~\bibnamefont{Kawahara}} \bibnamefont{{\it
  et~al}.}, \bibinfo{journal}{Phys. Rev. C} \textbf{\bibinfo{volume}{103}},
  \bibinfo{pages}{044001} (\bibinfo{year}{2021}).

\bibitem[{\citenamefont{Watanabe}(2020)}]{Atomu_PhD}
\bibinfo{author}{\bibfnamefont{A.}~\bibnamefont{Watanabe}}, Ph.D. thesis,
  \bibinfo{school}{Tohoku University} (\bibinfo{year}{2020}).

\bibitem[{\citenamefont{Nakai}(2021)}]{Nakai_dthesis}
\bibinfo{author}{\bibfnamefont{S.}~\bibnamefont{Nakai}}, Ph.D. thesis,
  \bibinfo{school}{Tohoku University} (\bibinfo{year}{2021}).

\bibitem[{\citenamefont{Dolph {\it et~al}.}(2011)\citenamefont{Dolph, Singh,
  Averett, Kelleher, Mooney, Nelyubin, Tobias, Wojtsekhowski, and
  Cates}}]{PhysRevC.84.065201}
\bibinfo{author}{\bibfnamefont{P.~A.~M.} \bibnamefont{Dolph}},
  \bibinfo{author}{\bibfnamefont{J.}~\bibnamefont{Singh}},
  \bibinfo{author}{\bibfnamefont{T.}~\bibnamefont{Averett}},
  \bibinfo{author}{\bibfnamefont{A.}~\bibnamefont{Kelleher}},
  \bibinfo{author}{\bibfnamefont{K.~E.} \bibnamefont{Mooney}},
  \bibinfo{author}{\bibfnamefont{V.}~\bibnamefont{Nelyubin}},
  \bibinfo{author}{\bibfnamefont{W.~A.} \bibnamefont{Tobias}},
  \bibinfo{author}{\bibfnamefont{B.}~\bibnamefont{Wojtsekhowski}},
  \bibnamefont{and} \bibinfo{author}{\bibfnamefont{G.~D.} \bibnamefont{Cates}},
  \bibinfo{journal}{Phys. Rev. C} \textbf{\bibinfo{volume}{84}},
  \bibinfo{pages}{065201} (\bibinfo{year}{2011}).

\bibitem[{\citenamefont{Romalis and Cates}(1998)}]{PhysRevA.58.3004}
\bibinfo{author}{\bibfnamefont{M.~V.} \bibnamefont{Romalis}} \bibnamefont{and}
  \bibinfo{author}{\bibfnamefont{G.~D.} \bibnamefont{Cates}},
  \bibinfo{journal}{Phys. Rev. A} \textbf{\bibinfo{volume}{58}},
  \bibinfo{pages}{3004} (\bibinfo{year}{1998}).

\bibitem[{\citenamefont{Coulter {\it et~al}.}(1988)\citenamefont{Coulter,
  McDonald, Happer, Chupp, and Wagshul}}]{COULTER198890}
\bibinfo{author}{\bibfnamefont{K.}~\bibnamefont{Coulter}},
  \bibinfo{author}{\bibfnamefont{A.}~\bibnamefont{McDonald}},
  \bibinfo{author}{\bibfnamefont{W.}~\bibnamefont{Happer}},
  \bibinfo{author}{\bibfnamefont{T.}~\bibnamefont{Chupp}}, \bibnamefont{and}
  \bibinfo{author}{\bibfnamefont{M.}~\bibnamefont{Wagshul}},
  \bibinfo{journal}{Nucl. Instr. and Meth. A} \textbf{\bibinfo{volume}{270}},
  \bibinfo{pages}{90} (\bibinfo{year}{1988}).

\bibitem[{\citenamefont{Alfimenkov {\it et~al}.}(1977)\citenamefont{Alfimenkov,
  Akopyan, Wierzbicki, Govorov, Pikelner, and Sharapov}}]{SovJNuclPhys.25.607}
\bibinfo{author}{\bibfnamefont{V.~P.} \bibnamefont{Alfimenkov}},
  \bibinfo{author}{\bibfnamefont{G.~G.} \bibnamefont{Akopyan}},
  \bibinfo{author}{\bibfnamefont{J.}~\bibnamefont{Wierzbicki}},
  \bibinfo{author}{\bibfnamefont{A.~M.} \bibnamefont{Govorov}},
  \bibinfo{author}{\bibfnamefont{L.~B.} \bibnamefont{Pikelner}},
  \bibnamefont{and} \bibinfo{author}{\bibfnamefont{E.~I.}
  \bibnamefont{Sharapov}}, \bibinfo{journal}{Sov. J. Nucl. Phys.}
  \textbf{\bibinfo{volume}{25}}, \bibinfo{pages}{607} (\bibinfo{year}{1977}).

\bibitem[{\citenamefont{Keith {\it et~al}.}(2004)\citenamefont{Keith,
  Chowdhuri, Rich, Snow, Bowman, Penttil\"a, Smith, Leuschner, Pomeroy, Jones
  {\it et~al}.}}]{PhysRevC.69.034005}
\bibinfo{author}{\bibfnamefont{C.~D.} \bibnamefont{Keith}},
  \bibinfo{author}{\bibfnamefont{Z.}~\bibnamefont{Chowdhuri}},
  \bibinfo{author}{\bibfnamefont{D.~R.} \bibnamefont{Rich}},
  \bibinfo{author}{\bibfnamefont{W.~M.} \bibnamefont{Snow}},
  \bibinfo{author}{\bibfnamefont{J.~D.} \bibnamefont{Bowman}},
  \bibinfo{author}{\bibfnamefont{S.~L.} \bibnamefont{Penttil\"a}},
  \bibinfo{author}{\bibfnamefont{D.~A.} \bibnamefont{Smith}},
  \bibinfo{author}{\bibfnamefont{M.~B.} \bibnamefont{Leuschner}},
  \bibinfo{author}{\bibfnamefont{V.~R.} \bibnamefont{Pomeroy}},
  \bibinfo{author}{\bibfnamefont{G.~L.} \bibnamefont{Jones}} \bibnamefont{{\it
  et~al}.}, \bibinfo{journal}{Phys. Rev. C} \textbf{\bibinfo{volume}{69}},
  \bibinfo{pages}{034005} (\bibinfo{year}{2004}).

\bibitem[{\citenamefont{Bouchiat {\it et~al}.}(1960)\citenamefont{Bouchiat,
  Carver, and Varnum}}]{PhysRevLett.5.373}
\bibinfo{author}{\bibfnamefont{M.~A.} \bibnamefont{Bouchiat}},
  \bibinfo{author}{\bibfnamefont{T.~R.} \bibnamefont{Carver}},
  \bibnamefont{and} \bibinfo{author}{\bibfnamefont{C.~M.}
  \bibnamefont{Varnum}}, \bibinfo{journal}{Phys. Rev. Lett.}
  \textbf{\bibinfo{volume}{5}}, \bibinfo{pages}{373} (\bibinfo{year}{1960}).

\bibitem[{\citenamefont{Babcock {\it et~al}.}(2003)\citenamefont{Babcock,
  Nelson, Kadlecek, Driehuys, Anderson, Hersman, and
  Walker}}]{PhysRevLett.91.123003}
\bibinfo{author}{\bibfnamefont{E.}~\bibnamefont{Babcock}},
  \bibinfo{author}{\bibfnamefont{I.}~\bibnamefont{Nelson}},
  \bibinfo{author}{\bibfnamefont{S.}~\bibnamefont{Kadlecek}},
  \bibinfo{author}{\bibfnamefont{B.}~\bibnamefont{Driehuys}},
  \bibinfo{author}{\bibfnamefont{L.~W.} \bibnamefont{Anderson}},
  \bibinfo{author}{\bibfnamefont{F.~W.} \bibnamefont{Hersman}},
  \bibnamefont{and} \bibinfo{author}{\bibfnamefont{T.~G.}
  \bibnamefont{Walker}}, \bibinfo{journal}{Phys. Rev. Lett.}
  \textbf{\bibinfo{volume}{91}}, \bibinfo{pages}{123003}
  (\bibinfo{year}{2003}).

\bibitem[{\citenamefont{Coulter {\it et~al}.}(1989)\citenamefont{Coulter,
  McDonald, Cates, Happer, and Chupp}}]{COULTER198929}
\bibinfo{author}{\bibfnamefont{K.}~\bibnamefont{Coulter}},
  \bibinfo{author}{\bibfnamefont{A.}~\bibnamefont{McDonald}},
  \bibinfo{author}{\bibfnamefont{G.}~\bibnamefont{Cates}},
  \bibinfo{author}{\bibfnamefont{W.}~\bibnamefont{Happer}}, \bibnamefont{and}
  \bibinfo{author}{\bibfnamefont{T.}~\bibnamefont{Chupp}},
  \bibinfo{journal}{Nucl. Instrum. Methods in Phys. Res. A}
  \textbf{\bibinfo{volume}{276}}, \bibinfo{pages}{29 } (\bibinfo{year}{1989}).

\bibitem[{\citenamefont{Gentile {\it et~al}.}(2017)\citenamefont{Gentile,
  Nacher, Saam, and Walker}}]{RevModPhys.89.045004}
\bibinfo{author}{\bibfnamefont{T.~R.} \bibnamefont{Gentile}},
  \bibinfo{author}{\bibfnamefont{P.~J.} \bibnamefont{Nacher}},
  \bibinfo{author}{\bibfnamefont{B.}~\bibnamefont{Saam}}, \bibnamefont{and}
  \bibinfo{author}{\bibfnamefont{T.~G.} \bibnamefont{Walker}},
  \bibinfo{journal}{Rev. Mod. Phys.} \textbf{\bibinfo{volume}{89}},
  \bibinfo{pages}{045004} (\bibinfo{year}{2017}).

\bibitem[{Coh()}]{Coherent-DILAS}
\bibinfo{note}{Coherent-DILAS, E11.4B-795.2-75C-SO3.6(1x1)}.

\bibitem[{\citenamefont{Chen {\it et~al}.}(2014)\citenamefont{Chen, Gentile,
  Ye, Walker, and Babcock}}]{JApplPhys.116.014903}
\bibinfo{author}{\bibfnamefont{W.~C.} \bibnamefont{Chen}},
  \bibinfo{author}{\bibfnamefont{T.~R.} \bibnamefont{Gentile}},
  \bibinfo{author}{\bibfnamefont{Q.}~\bibnamefont{Ye}},
  \bibinfo{author}{\bibfnamefont{T.~G.} \bibnamefont{Walker}},
  \bibnamefont{and} \bibinfo{author}{\bibfnamefont{E.}~\bibnamefont{Babcock}},
  \bibinfo{journal}{J. Appl. Phys.} \textbf{\bibinfo{volume}{116}},
  \bibinfo{pages}{014903} (\bibinfo{year}{2014}).

\bibitem[{\citenamefont{Ino}(2019)}]{Ino_2019}
\bibinfo{author}{\bibfnamefont{T.}~\bibnamefont{Ino}}, \bibinfo{journal}{J.
  Phys.: Conf. Ser.} \textbf{\bibinfo{volume}{1316}}, \bibinfo{pages}{012012}
  (\bibinfo{year}{2019}).

\bibitem[{\citenamefont{Ikeda {\it et~al}.}(2016)\citenamefont{Ikeda, Taketani,
  Takamura, Sunaga, Kumagai, Oba, Otake, and Suzuki}}]{IKEDA201661}
\bibinfo{author}{\bibfnamefont{Y.}~\bibnamefont{Ikeda}},
  \bibinfo{author}{\bibfnamefont{A.}~\bibnamefont{Taketani}},
  \bibinfo{author}{\bibfnamefont{M.}~\bibnamefont{Takamura}},
  \bibinfo{author}{\bibfnamefont{H.}~\bibnamefont{Sunaga}},
  \bibinfo{author}{\bibfnamefont{M.}~\bibnamefont{Kumagai}},
  \bibinfo{author}{\bibfnamefont{Y.}~\bibnamefont{Oba}},
  \bibinfo{author}{\bibfnamefont{Y.}~\bibnamefont{Otake}}, \bibnamefont{and}
  \bibinfo{author}{\bibfnamefont{H.}~\bibnamefont{Suzuki}},
  \bibinfo{journal}{Nucl. Instr. and Meth. A} \textbf{\bibinfo{volume}{833}},
  \bibinfo{pages}{61} (\bibinfo{year}{2016}).

\bibitem[{\citenamefont{Hirota {\it et~al}.}(2005)\citenamefont{Hirota,
  Shinohara, Ikeda, Mishima, Adachi, Morishima, Satoh, Oku, Yamada, Sasao {\it
  et~al}.}}]{PCCP.7.1836}
\bibinfo{author}{\bibfnamefont{K.}~\bibnamefont{Hirota}},
  \bibinfo{author}{\bibfnamefont{T.}~\bibnamefont{Shinohara}},
  \bibinfo{author}{\bibfnamefont{K.}~\bibnamefont{Ikeda}},
  \bibinfo{author}{\bibfnamefont{K.}~\bibnamefont{Mishima}},
  \bibinfo{author}{\bibfnamefont{T.}~\bibnamefont{Adachi}},
  \bibinfo{author}{\bibfnamefont{T.}~\bibnamefont{Morishima}},
  \bibinfo{author}{\bibfnamefont{S.}~\bibnamefont{Satoh}},
  \bibinfo{author}{\bibfnamefont{T.}~\bibnamefont{Oku}},
  \bibinfo{author}{\bibfnamefont{S.}~\bibnamefont{Yamada}},
  \bibinfo{author}{\bibfnamefont{H.}~\bibnamefont{Sasao}} \bibnamefont{{\it
  et~al}.}, \bibinfo{journal}{Phys. Chem. Chem. Phys.}
  \textbf{\bibinfo{volume}{7}}, \bibinfo{pages}{1836} (\bibinfo{year}{2005}).

\bibitem[{\citenamefont{Normand {\it et~al}.}(2016)\citenamefont{Normand,
  Jiang, Brown, Robertson, Crow, and Tong}}]{Normand_2016}
\bibinfo{author}{\bibfnamefont{E.}~\bibnamefont{Normand}},
  \bibinfo{author}{\bibfnamefont{C.~Y.} \bibnamefont{Jiang}},
  \bibinfo{author}{\bibfnamefont{D.~R.} \bibnamefont{Brown}},
  \bibinfo{author}{\bibfnamefont{L.}~\bibnamefont{Robertson}},
  \bibinfo{author}{\bibfnamefont{L.}~\bibnamefont{Crow}}, \bibnamefont{and}
  \bibinfo{author}{\bibfnamefont{X.}~\bibnamefont{Tong}},
  \bibinfo{journal}{Journal of Physics: Conference Series}
  \textbf{\bibinfo{volume}{711}}, \bibinfo{pages}{012012}
  (\bibinfo{year}{2016}).

\end{thebibliography}

\end{document}